\documentclass[runningheads]{llncs}
\usepackage{graphicx,amssymb,amsmath,mathtools,wasysym,floatrow}
\usepackage[outercaption]{sidecap}    
\usepackage[ruled,vlined,linesnumbered]{algorithm2e}
\usepackage[dvipsnames]{xcolor}

\DeclareMathOperator*{\argmax}{arg\,max}

\DeclareMathSymbol{\mlq}{\mathord}{operators}{``}
\DeclareMathSymbol{\mrq}{\mathord}{operators}{`'}
\newcommand\EP{\mathit{EP}}
\newcommand\AT{\mathit{AT}}
\newcommand\ssf{\mathit{ssf}}
\newcommand\sitf{\mathit{sf}}
\newcommand\fg{\smiley{}}
\newcommand\pg{\frownie{}}
\newcommand\csf{\mathit{csf}}

\newcommand\Ss{\mathit{SS}}
\newif\ifComments
\Commentstrue
\ifComments
\newcommand{\mdb}[1]{\textcolor{blue}{Mark: #1}}
\newcommand{\mm}[1]{\textcolor{magenta}{Mehran: #1}}
\newcommand{\maa}[1]{\textcolor{red}{Mohammad: #1}}
\newcommand{\msq}[1]{\textcolor{green}{Mahnaz: #1}}
\else
\newcommand{\mdb}[1]{}
\newcommand{\mm}[1]{}
\newcommand{\maa}[1]{}
\newcommand{\msq}[1]{}
\fi

\begin{document}
	\title{Fairness-Aware Process Mining}
	\author{Mahnaz Sadat Qafari \and
		Wil van der Aalst}
	\authorrunning{M. S. Qafari et al.}
	% First names are abbreviated in the running head.
	% If there are more than two authors, 'et al.' is used.
	%
	\institute{Rheinisch-Westfälische Technische Hochschule Aachen(RWTH), Aachen, Germany \\
		\email{m.s.qafari@pads.rwth-aachen.de,wvdaalst@pads.rwth-aachen.de}}
	
	\maketitle 
	\begin{abstract}
	Process mining is a multi-purpose tool enabling organizations to improve their processes. One of the primary purposes of process mining is finding the root causes of performance or compliance problems in processes. The usual way of doing so is by gathering data from the process event log and other sources and then applying some data mining and machine learning techniques. However, the results of applying such techniques are not always acceptable. In many situations, this approach is prone to making obvious or unfair diagnoses and applying them may result in conclusions that are unsurprising or even discriminating (e.g., blaming overloaded employees for delays). In this paper, we present a solution to this problem by creating a fair classifier for such situations. The undesired effects are removed at the expense of reduction on the accuracy of the resulting classifier. We have implemented this method as a plug-in in ProM. Using the implemented plug-in on two real event logs, we decreased the discrimination caused by the classifier, while losing a small fraction of its accuracy.
	\end{abstract}
	%---------------------------------------------------------
	\section{Introduction}
    \paragraph{Motivation.}    
Academic and commercial process mining tools aim to find the root causes of performance or compliance problems in processes. Mainly, a classifier, say a decision tree, is created using the data gathered from the process and then the rule mining is done using that decision tree~\cite{de2016general}. However, this approach may lead to diagnoses that are not valuable. In some cases, the main cause of the problem is already known and essentially cannot be altered. Also, due to the strong correlation of the known main cause and the problem, it may become impossible to see the other minor but probably more practically valuable causes of the problem. % In other cases, the consequence of these diagnoses might be false and unfair accusations, blaming a group of employees or customers.
 Consider the following two scenarios: 
\begin{itemize}
	\item [(i)] there is a bottleneck in the process and it is caused by the most busy employee, or 
	\item [(ii)] there are deviations caused by the most experienced resources taking the most difficult cases.
\end{itemize}
In these scenarios, it is likely that the most busy employees or the most experienced resources are declared the main reasons for the bottleneck or deviations in the process. This is not just unfair but also does not provide novel insights (just stating the obvious). Even if we remove the attribute conveying the employee or the resource, still rules that proxy these attributes would be revealed as the result of the traditional rule mining~\cite{zemel2013learning}. In these cases, it is essential to make inference about the less trivial root-causes of the problem in the process.%In many cases, such diagnoses are addressing those issues in the process that can not be altered or improved. 

As another application, consider that for a given process we are interested in questions which are related to investigating the process while ignoring the effect of different values of a particular attribute. ``Following the progress of career paths while eliminating gender differences" is one example of these sorts of situations where we need to remove the correlation between two attributes in the data. Here we need to infer the rules governing the process ignoring the correlation between the effect of different values of that particular attribute and the process attribute of interest.

% while respecting fairness. 

\paragraph{Discrimination-aware data mining.} Each population can be partitioned into several subgroups according to the properties of its members, e.g., race, age, or academic degree. \emph{Discrimination} means treating a subgroup of people, called \emph{sensitive group}, in an unfair way merely because of being a member of that subgroup. There is a possibility that negligent usage of new advanced technologies, especially in the field of data mining and machine learning, inadvertently cause discrimination. To avoid these phenomena, % unpremeditated discriminatory results by the technology, 
detecting discrimination and designing fair predictors have been studied intensively(see~\cite{bolukbasi2016man,hardt2016equality,kleinberg2016inherent,kusner2017counterfactual,zemel2013learning,zliobaite2015survey}). This paper is the first paper to address this problem in the context of process mining.
	
	\emph{Demographic parity} is one of the most studied definitions of fairness that indicates the portion of people in the sensitive subgroup who receive the desired result must be the same as the whole population. To maintain this criterion, in some approaches the training data is manipulated. This is done by changing the labels of some items in  the training data~\cite{kamiran2010classification,kamiran2012data,zemel2013learning}, or by applying carefully designed re-sampling methods~\cite{kamiran2012data}. In another approach,~\cite{zemel2013learning}, the representation of the data is changed, and the fairness is maintained as a side-effect of fair representations. In~\cite{kamiran2010discrimination}, demographic parity in a decision tree is retained by taking into account the information gain of the sensitive attribute as well as the class attribute as the criteria used for splitting the internal nodes. In~\cite{kamiran2010discrimination}, the relabeling technique is also used to modify leaves of the resulting decision tree. 

Besides demographic parity, other notions of fairness have been formalized in the literature. For example, \emph{equal opportunity} and \emph{equal odds} have been introduced in~\cite{Hardt:2016:EOS:3157382.3157469}, and \emph{individual fairness} has been introduced in~\cite{dwork2012fairness}. We refer the interested readers to~\cite{berk2018fairness} for a review of various fairness criteria.
%	\paragraph{Related work.} A \emph{classifier} is a function that assigns items in a collection to labels or classes. Usually, a classifier is trained using a set of labeled data items, and its goal is to accurately predict the class (label) for new items of the data. The behavior of a classifier is highly influenced by the properties of the data that have been used for its training. If there is discrimination in the training data, then there is a possibility that the trained classifier performs as a proxy of or even amplifies the discrimination. There is a notable amount of work in the area of data mining dedicated to designing algorithms that make \emph{fair} predictions; i.e., nondiscriminatory predictions (see~\cite{bolukbasi2016man,hardt2016equality,kleinberg2016inherent,kusner2017counterfactual,zemel2013learning,zliobaite2015survey}).	
		\paragraph{Process mining. }Process mining is the link between model-based process analysis and data-oriented analysis techniques; a set of techniques that support the analysis of business processes based on event logs. In this context, several works have been dedicated to decision mining and finding the correlation among the process data and making predictions~\cite{sani2017subgroup,leemans2014process,de2016general,rozinat2006decision}. 
		
		Ethical and legal effects of process mining can be considered in two categories; confidentiality and fairness issues. Confidentiality in the process mining has recently receiving attention~\cite{RafieiWA18}. To the best of our knowledge, there is no work in the area of process mining dedicated to investigating fairness issues. This is the first publication considering discrimination within a given process.      	
\paragraph{Our Results.}    
We provide a solution for the previously mentioned problems. Specifying a problem in the process, we propose an approach by adopting the techniques available in data mining for removing discrimination from classifiers in the area of process mining to avoid unfair or obvious conclusions in such scenarios. We do that by declaring the attribute that indicates the existence of the problem in the given situation as the \emph{class attribute} and the attribute that we want to decrease its dependency to the class attribute as the \emph{sensitive attribute}. We consider the class attribute to be binary with the following two values: $+$ indicates the desirable result conveying the problem of interest has not been faced while $-$ has the opposite meaning. The sensitive attribute is also assumed to be binary, where $\pg$ convey belonging to the sensitive group while $\fg$ convey belonging to the rest of the population (favorable group). In the previous examples, we can consider the existence of the delay or deviation in the cases as the class attribute and the employee being busy or highly experienced as the sensitive attribute. Now, we can consider the problem as a discriminatory case and remove the dependency of the class and the sensitive attributes in the resulting classifier by creating a fair classifier. Doing so, the resulting rules would not be discriminatory against the sensitive group. Also, this technique masks some of the causes of that problem and focus on the other ones. Consequently, we can aim at finding more feasible solutions for the problem in the process by finding other possibly less obvious or even hidden causes of the problem.

\begin{figure}\label{pic::snapshot}
	\includegraphics[width=120mm]{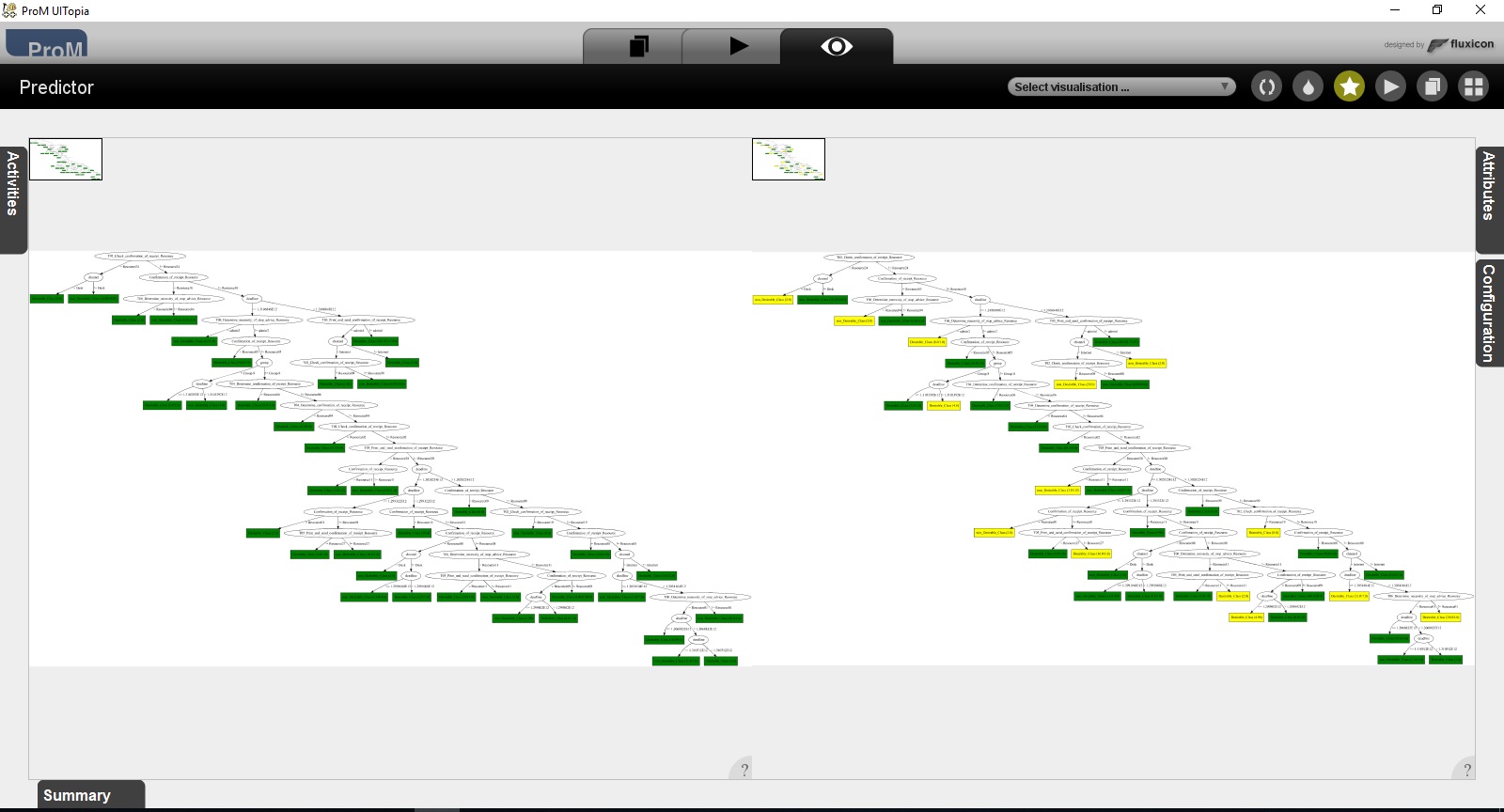}
	\caption{The two decision trees created by the implemented plug-in. The left one is a standard decision tree and the right one is a fair decision tree. The yellow leaves are the ones that have been relabeled to ensure fairness.}
\end{figure}

We have implemented this approach as a ProM plug-in. In this plug-in, we create two decision trees using the data gathered about the process. The first one is a standard decision tree (using algorithm C4.5) describing the correlations between dependent and independent attributes as they are. The second one is a fair decision tree, in which the dependency between the sensitive and the class attributes is decreased or removed. 
The goal is to create a decision tree with no unacceptable discrimination and the maximum possible accuracy. A snapshot of the plug-in implemented in ProM is depicted in Figure~\ref{pic::snapshot}.   

The remainder of the paper is organized as follow. In Section~\ref{problem}, we present the problem statement. A high-level overview of the proposed approach is presented in Section~\ref{approach}. The experimental results of applying the implemented method on some real event logs are presented in Section~\ref{result}. Finally, in Section~\ref{conclution}, we summarize our approach and discuss directions for further research.

\section{Problem Statement}\label{problem}
To analyze conformance and performance problems, we use event data and process models (discovered or hand-made).\footnote{We assume the reader to be familiar with the concepts like set, multi-set, and function. Given a non-empty set $X$, we denote all the non-empty subsets of $X$ by $\mathbb{P}(X)$. Given two sets $A$ and $B$, a partial function $f: A \not \mapsto B$ is defined as a function $f:A'\mapsto B$ for some $A'\subseteq A$. We say $f(a) = \bot$ if $a \not\in A'$.} An event log is a collection of traces and each trace is a collection of events related to the same case. Also each trace may be associated with some attributes. Consider $\mathcal{U}_{act}$ as the universe of all possible \emph{activity names},  $\mathcal{U}_{time}$ the universe of all possible \emph{time stamps}, $\mathcal{U}_{att}$ the universe of all possible \emph{attribute names}, $\mathcal{U}_{val}$ the universe of all possible \emph{values}, and, $\mathcal{U}_{map}:\mathcal{U}_{att} \not \mapsto \mathcal{U}_{val}$. Also, let $values: \mathcal{U}_{att} \mapsto \mathbb{P}(\mathcal{U}_{val})$ be the function that returns the set of all possible values for each attribute name. We define an event log as follow:
\begin{definition}[Event Log]\label{def:universe}
	An \emph{event} is an element of $\mathcal{U}_{act} \times \mathcal{U}_{time} \times \mathcal{U}_{map} $ and the universe of all possible \emph{events} is denoted by $\mathcal{E}$. A \emph{log} is an element of $\mathbb{P}(\mathcal{U}_{map}\times \mathbb{P}(\mathcal{E}))$ and the universe of all possible logs is denoted by $\mathcal{L}$. We call each $t\in L$, where $L \in \mathcal{L}$, a \emph{trace}. 
\end{definition}
%We assume that in a given event log $L \in \mathcal{L}$ each event has a unique time stamp.    
To work with event logs, we need the following helper functions:
\begin{itemize}
	\item Given an event $e = (act,time,map)\in \mathcal{E}$, $\pi_{act}(e) = act$, $\pi_{time}(e) = time$, and, $\pi_{map}(e) = map$.
	\item Given $t= (map, E) \in L$, where $L\in \mathcal{L}$, then $\pi_{map}(t) = map$ and $\pi_{events}(t) = E$.
	\item Given $E \in \mathbb{P}(\mathcal{E})$, then $\pi_{maxtime}(E) = \argmax_{e \in E} \pi_{time}(e)$, $\pi_{act}(E) = \{ e \in E | \pi_{act}(e) = act\}$, and, $E_{\leq time} =\{e \in E| \pi_{time}(e) \leq time\}$. This function returns the set of events of a trace whose time stamps are at most equal to a given time stamp.
\end{itemize}
We assume that each event in a given log $L$ is unique and also has a unique time stamp. In other words, $\forall t,t'\in L \forall e\in \pi_{events}(t) \forall e'\in \pi_{events}(t') \big( \pi_{time}(e) =\pi_{time}(e')\implies (e=e'\land t=t')\big)$.

If the problem in the process is about the traces, like delay in some cases, then for a given trace all the values of its trace and event level attributes might be relevant. However, if the problem is related to a specific activity, like a bottleneck in activity $act$, then we need to extract the data from the trace attributes plus the attributes of a subset of its events that occur before the occurrence of that specific event. Also, the class attribute may occur several times in a given trace. We define the notion of a \emph{situation} to handle such cases as follows: 
\begin{definition}[Situation]\label{def:situation}
	We define a \emph{situation} as an element in $(\mathcal{U}_{map}\times \mathbb{P}(\mathcal{E}))$. The set of all possible situations is denoted by $\mathcal{U}_{sit}$. Given a log $L\in \mathcal{L}$, we define the set of all situations derived from it as:
	$$S_L=\bigcup_{(map,E)\in L}\ \big( \bigcup_{e \in E}\ \ \ \{(map,E_{\leq \pi_{time}(e)})\} \big).$$
	It is obvious that $L \subseteq S_L$. Any $S \subseteq S_L$ is called a \emph{situation subset} of $L$. For a given log $L$, there are two main types of situation subsets. The first one is the \emph{trace situation subset} which is $S_{L,\bot} = L$. The second type is the \emph{event specified situation subsets} which includes all $S_{L,act} =\{ (map , E)\in S_L| \pi_{act}(\pi_{maxtime}(E))=act\}$, where $act\in \mathcal{U}_{act}$ and $S_{L,act} \neq \emptyset$.    
\end{definition}

%Informally, given a log $L$ and a trace $(map, E)\in L$, this trace can be mapped to several situations which each one of them is a sub-trace including all the 

An attribute may be a trace or an event level attribute. So, it may happen several times with different values in a trace. To specify an attribute, besides the name of the attribute, we need to know if it is a trace or an event attribute and if it is an event attribute, we need to know to which events does it belong. To concretely specify an attribute, we use the \emph{situation feature} notion defined as follows:
\begin{definition}[Situation Feature]\label{def:sf}
	For any given $ a \in \mathcal{U}_{act}\cup \{\bot \}$ and $att \in \mathcal{U}_{att}$, we call $ \sitf_{a,att}:\mathcal{U}_{sit}\not \mapsto \mathcal{U}_{val}$ a \emph{situation feature}. Given a situation $(map, E)$ we define $\sitf_{a,att}((map,E))$ as follows:
	$$\sitf_{a,att}((map,E)) = \begin{cases} map(att) & a = \bot\\ {\pi_{map}}(\pi_{maxtime}(\pi_{a}(E)))(att) & a \in \mathcal{U}_{act}\end{cases}.$$
	We denote the universe of all possible situation features by $\mathcal{U}_{\sitf}$.	Given a situation feature $\sitf_{a,att}$, we define $values(\sitf_{a,att})=values(att)$. Also, for a given $n \in \mathbb{N}$, $\EP \in \mathcal{U}_{\sitf}^n$ is a \emph{situation feature extraction plan} of size~$n$, where $\mathcal{U}_{\sitf}^n$ is defined as $ \underbrace{\mathcal{U}_{\sitf} \times \dots \times \mathcal{U}_{\sitf}}_{n\ times}$.
\end{definition}
A situation feature extraction plan can be interpreted as the schema, the tuple composed of those situation features that are relevant to the given problem in the process.

The first step of solving any problem is concretely specifying the problem. We call such a problem description a \emph{situation specification} which is defined as follows:
\begin{definition}[Situation Specification]\label{def::ss}
	A situation specification is a tuple $\Ss =(\EP,\ssf,\csf,\epsilon)$ in which
	\begin{itemize}
		\item [(i)] $\EP \in \mathcal{U}_{\sitf}^n$, where $n \in \mathbb{N}$, is the situation feature extraction plan which includes all the situation features for which we are going to investigate their effect on the given problem.
		\item [(ii)] $\ssf \in \mathcal{U}_{\sitf}$, the sensitive situation feature where $values(\ssf) = \{\fg,\pg \}$ and $\ssf \not\in \EP$.
		\item [(iii)] $\csf \in \mathcal{U}_{\sitf}$, the class situation feature where $values(\csf) = \{+,-\}$, $\csf \not\in \EP $, and $\csf \neq \ssf$. 
		\item [(vi)] $\epsilon \in [0,1]$, indicating the acceptable level of discrimination against $\ssf$ (the amount of acceptable dependency between $\ssf$ and $\csf$).
	\end{itemize}
\end{definition}
%For a given situation $s$, the value of $\ssf (s)$ indicates if the $s$ belong to the sensitive group or not. Also 
%In a given situation specification, $\epsilon$ 

For a given situation specification, we go through the following three steps;
	\begin{enumerate}
		\item \textbf{Enriching the log:} The event log is enriched with several attributes that are driven from the given event log or other sources of information.
		\item \textbf{Extracting the data:} The relevant independent situation features, sensitive and class situation features values are driven from the enriched log.
		\item \textbf{Learning fair classifier:} Two decision tree classifiers are created by the C4.5 algorithm% implemented as J48 tree in the WEKA package
		. Then a relabeling technique is used to remove the unacceptable discrimination from one of the decision trees. 
	\end{enumerate}
	\begin{figure}\label{pic::general}
		\includegraphics[width=130mm]{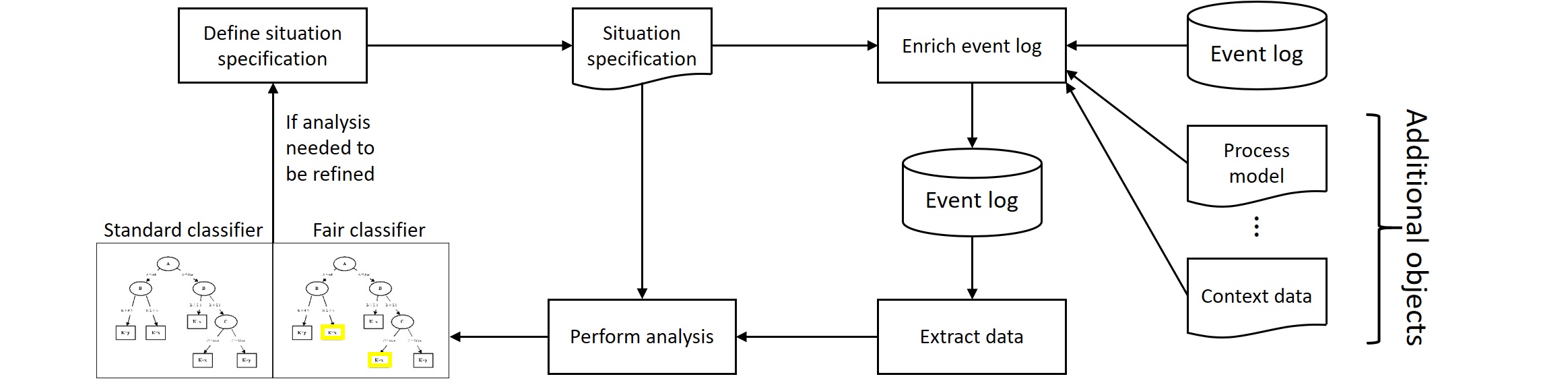}
		\caption{The general framework proposed for fair root-cause analysis. First, according to the situation specification the event log is enriched by preprocessing the log and other sources of information. Then, the data is extracted from the enriched event log. Finally, two standard and fair classifier are created. Based on the analysis result, it is possible to adapt the situation specification to gather additional insights.}
	\end{figure}
	%The details of these three steps are presented in the next section. 
	The general approach of our method is depicted in Figure~\ref{pic::general}. In this approach, the first two steps are related to the data extraction and the third one aims at creating a fair classifier.

	\section{Approach}\label{approach}
	We go through every one of the steps for creating a fair classifier for a given situation specification, mentioned in the previous section, in more details.
	\paragraph{1. Enriching the log.} Let $\Ss =(\EP,\ssf,\csf, \epsilon)$ be the given situation specification. If $\EP$ includes situation features that can not be directly extracted from the given log $L$, we enrich the log by augmenting each trace of it. %The formal definition of trace augmentation is presented as definition \ref{def:aug}. Loosely speaking, 
	In this step, we add some attribute values to the traces and its events. These added attributes can be related to any one of different process characteristics; time perspective, data flow-perspective, control-flow perspective, conformance perspective, or resource organization perspective. They may be driven from the given log, conformance checking results from replaying the traces on a given Petri-net model, or any external information resource like the weather information. Here, we assume that we have access to the log, Petri-net modeling the process of the log and the conformance checking results of replaying the log on the given model. The range of possible attributes that can be used to enrich the log is broad. We have implemented some of them. For example, some of the implemented trace attributes are \emph{trace-duration}, \emph{trace-delay}, \emph{sub-model-duration }, and, \emph{deviation}. Some of the implemented event attributes are \emph{next-activity-in-the-trace}, \emph{previous-activity-in-the-trace}, \emph{total-workload}, and \emph{resource-workload}. %\footnote{Total workload indicates the number of events under execution in the process at the time that the given event occurs.}, and, \emph{Resource workload}\footnote{Resource workload indicates the number of events under execution by the same resource that is executing the given event at the time it occurs.}. 
%
	 %It is worth noting that these are just some of the possible attributes that have been implemented in the ProM plug-in.
	 The formal definitions of these attributes are beyond the scope of this paper\footnote{Refer to~\cite{de2016general} for the formal definition of some of the properties.}.
	\paragraph{Extracting the data.} To discover meaningful dependency results by the decision tree, we need to capture the data such that the causality relations among them and the class attribute are not violated. To do so, given  $\csf =\sitf_{a,att}$, we apply the following two rules while extracting the data;
	\begin{enumerate}
		\item If $a \in \mathcal{U}_{act}$, each trace may map to several situations and the data should be extracted from that part of the trace that happens before the occurrence of $\csf$. If $a =\bot$, then $\csf$ is related to a trace level attribute and the data should be extracted from the whole trace.
		\item The value of the independent situation feature with the closest occurrence time to the occurrence of $\csf$ must be collected.
	\end{enumerate}
	The second rule is valid assuming that if one of the independent situation features has happened several times before the occurrence of $\csf$ in a trace, the one that is closest to the occurrence of $\csf$ regarding the time, has the most effect on its value.
	
	To follow the first rule, for the given log $L$ and the situation specification $\Ss=(\EP,\ssf,\csf,\epsilon)$, where $\csf =\sitf_{a,att}$, we set $S=S_{L,\bot}$ if $a=\bot$ and we set $S=S_{L,act}$ if $a = act$. If $S=S_{L,\bot}$, then each situation is a trace in the log. However, if $S=S_{L,act}$, then for all $s \in S$ we have $\pi_{maxtime}(\pi_{events}(s)) = act$. i.e., $S$ is the set of all situations in $S_L$ for which the activity name of the event with maximum time stamp is $act$.
	
	The final step for extracting the data is creating a data table and annotating each row of the table by adding the values of sensitive and class situation feature to it. We create the data table and annotate it as specified in the following definition. Note that each situation is mapped to a row of the table.
		\begin{definition}[Situation Feature Table]	\label{def:table}
		Given a situation feature extraction plan $\EP=(\sitf_{a_1,att_1}, \dots , \sitf_{a_n,att_n})$, and a situation set $S \subseteq \mathcal{U}_{sit}$, a \emph{situation feature table} is a multi-set which is defined as:
		$$T_{S,\EP}=[ ( \sitf_{a_1,att_1}(s),\dots , \sitf_{a_n,att_n}(s) )| s \in S].$$ 
		For a log $L\in \mathcal{L}$, $S\subseteq S_L$, we call $T_{S,\EP}$ a \emph{situation feature table of $L$}. 
		%	\end{definition}
		%	\begin{definition}[Annotated Situation Feature Table]
		
		For a given situation feature table $T_{S,\EP}$ and $\csf,\ssf \in \mathcal{U}_{\ssf}$ for which $\ssf \neq \csf$ and $\csf,\ssf \not\in \EP$ and $\forall_{s \in S}\ (\csf(s)\neq \bot \land \ssf(s) \neq \bot) $, we define an \emph{annotated situation table} $\AT_{S,\EP,\ssf,\csf}$ as:		
		$$\AT_{S,\EP,\ssf,\csf}=[( \sitf_{a_1,att_1}(s),\dots , \sitf_{a_n,att_n}(s), \ssf(s), \csf(s) ) | s \in S].$$
		We call each element of $\AT_{S,\EP,\ssf,\csf}$ an \emph{instance}. 
		For a given instance  $inst_s= ( \sitf_{a_1,att_1}(s), \dots , \sitf_{a_n,att_n}(s) , \ssf(s), \csf(s))$ we define $\pi_{\EP} (inst_s)=\big(\sitf_{a_1,att_1}(s),\\ \dots , \sitf_{a_n,att_n}(s) \big)$, $\pi_{\ssf}(inst_s)=\ssf(s)$, and, $\pi_{\csf}(inst_s)=\csf(s)$.
	\end{definition}
	Here, $\ssf$ is the sensitive and $\csf$ is the class (label) situation feature. Also, each member of $inst_s \in \AT_{S,\EP,\ssf,\csf}$ where $s \in S$
	can be seen as a row of the data table in which $\pi_{\EP}(inst_s)$ is the tuple including independent attribute values and $\pi_{\csf}(inst_s)$ is the class attribute value of $inst_s$.
	\paragraph{Learning fair classifier.}	
	We define a classifier as follow:
	\begin{definition}[Classifier]\label{def:classifier}
		Let $S$ be a set of situations and $\EP= ( \sitf_{a_1,att_1}, \dots ,\\ \sitf_{a_n,att_n})$ be a situation extraction plan and $\csf\in \mathcal{U}_{\sitf}$ such that $\forall_{1\leq i \leq n} \sitf_{a_i,att_i} \neq \csf$, then a \emph{classifier} is a function $class:T_{S,\EP} \mapsto values(\csf)$.	
	\end{definition}
Given a classifier $class$ and an annotated situation table $\AT_{S,\EP,\ssf,\csf}$, then the accuracy of $class$ over $\AT_{S,\EP,\ssf,\csf}$ is measured as:
%\begin{equation}\label{accuracy}
$$acc(class,\AT_{S,\EP,\ssf,\csf}) = \frac{|[inst \in \AT_{S,\EP,\ssf,\csf}| class(\pi_{\EP}(inst))=\pi_{\csf}(inst)]|}{|\AT_{S,\EP,\ssf,\csf}|}.$$
%%\end{equation}
	For fairness, we use demographic parity as the main concept. To measure the discrimination in the data, we use the measure mentioned in~\cite{kamiran2010discrimination}, which is:
%	\begin{equation}\label{fairness}
	$$disc(\AT_{S,\EP,\ssf,\csf})= \frac{|[inst \in \AT_{S,\EP , \ssf , \csf} |\pi_{\ssf}(inst)=\fg \land \pi_{\csf}(inst)=+]|}{|[inst \in \AT_{S,\EP , \ssf , \csf}| \pi_{\ssf}(inst)=\fg]|}- $$ 
	$$\frac{|[inst \in \AT_{S,\EP , \ssf, \csf}| \pi_{\ssf}(inst)=\pg \land \pi_{\csf}(inst)=+)]|}{|[inst \in \AT_{S,\EP , \ssf , \csf}| \pi_{\ssf}(inst)=\pg]|}.$$
	%\end{equation}
	%defined as $\AT_{S,\EP , \ssf , \csf}$ in
   % This measure simply is the portion of the people who are favorable and received the desirable outcome minus the portion of people in the protected group who received the desirable outcome. 
   By replacing $\pi_{\csf}(inst)$ with $class(\pi_{\EP}(inst))$ in this equation, we can measure the discrimination imposed by the classifier $class$.

For the classifier, we use decision trees. It is worth mentioning that both the classifier and the measure of discrimination can be changed according to the given application. 

The first step toward removing the discrimination has already been taken during the creation of the classifier by not considering the sensitive situation feature for the classification purpose (Definition~\ref{def:classifier}). As mentioned in many works, e.g.~\cite{zemel2013learning}, this is not enough due to the existence of correlation among different situation feature values in a given situation feature table. The discrimination in the classifier can be further eliminated by the relabeling technique. In this paper, we relabel leaves in the decision tree to balance accuracy and fairness. However, other discrimination free classifiers can be used~\cite{aghaei2019learning,zafar2017fairness,zemel2013learning}.

As mentioned before, in the implemented plug-in two classifiers are generated. The first one is a tree classifier that is generated by J48 tree classifier implementation of C4.5 algorithm in WEKA package. Then, if the discrimination in the resulting decision tree is more than an acceptable threshold $\epsilon$, the leaves of the decision tree are relabeled to create a fair classifier. For the relabeling, we use an algorithm similar to the one mentioned in~\cite{kamiran2010discrimination}. In~\cite{kamiran2010discrimination}, the leaves of the tree are ordered in descending order of the ratio of the discrimination gain and accuracy lose of relabeling each leaf. Then according to this order, leaves are relabeled until the discrimination in the classifier tree is lower than $\epsilon$. As mentioned in~\cite{kamiran2010discrimination}, the problem of finding the set of leaves to be relabeled such that the discrimination in the decision tree is lower than a given threshold $\epsilon$ with the lowest possible negative effect on the accuracy of the decision tree is equivalent to the knapsack problem. In the relabeling algorithm implemented in the ProM plug-in, we use dynamic programming and rounding to choose approximately the best possible set of leaves to be relabeled.

Note that in the context of process mining and root cause analysis, changing the class label from + to - and from - to + at the same time may not be desirable. Consider a case where the problem in the process is the delay in some cases. Then + label of a situation means being on-time. Changing the label of some leaves from + to - means considering some delay where they do not exist. So in some cases we may need to restrict the relabeling technique to just desirable or just undesirable labeled leaves of the tree. Note that if we restrict the relabeling, there might be cases where the discrimination of the fair tree is higher than given $\epsilon$. In these cases the discrimination in the fair decision tree would be close $\epsilon$.
	
		\section{Implementation and Experimental Results}\label{result}
		The approach presented in Section \ref{approach} has been implemented as a plug-in of ProM which is an open source framework for process mining. The implemented plug-in is available under the name \emph{Discrimination aware decision tree}.
		
		The inputs of the plug-in are the event log, the Petri-net model of the process, and, the conformance checking results of replaying the given log on the given model. Using these inputs, several derivative attributes for enriching traces have been implemented. 
		
		The current implementation focuses on three types of problems in a given process: %The set of augmentation and the set of independent attributes		
		%Here we target three different types of problems regarding a process; 
		\begin{itemize}
			\item \textbf{Routing problems: }When there is a choice in the model of the process, what was the reason that some cases took one choice while the others took another?
			\item \textbf{Deviation problems: }This category refers to the questions like why a specific activity has been skipped or what was the reason that some traces do not comply with the given model.
			\item \textbf{Performance: }This category includes questions like what was the reason for the delay in the cases or why a specific activity takes more time in some cases.
		\end{itemize}
		To illustrate the fair analysis of these problems we use two real data logs, the \emph{hospital billing}\footnote{https://data.4tu.nl/repository/collection:event\_logs\_real} event log and \emph{receipt phase of an environmental permit application process (WABO) CoSeLoG project}\footnote{https://data.4tu.nl/repository/uuid:a07386a5-7be3-4367-9535-70bc9e77dbe6} (receipt log for short). We use the last 20000 traces of hospital billing log in the experiments which include 71188 activities. The receipt log includes 1434 traces and 8577 activities. In this initial evaluation, we created a controlled experiment with a known ground truth. The discrimination is added to the event logs artificially and then the altered logs are used to evaluate the method and investigate the effect of removing discrimination on the accuracy of the created fair decision trees. In all the experiments the same setting has been used. For example in all the experiments $\epsilon =0.05$ and there was no limit for applying relabeling technique. Also for each event log, the same set of independent situation features has been chosen and all the parameters for creating the decision tree were the same. 60 percent of the data has been used for training, and 40 percent of the data has been used for testing the classifier. 
	\begin{figure}
		\includegraphics[width=120mm]{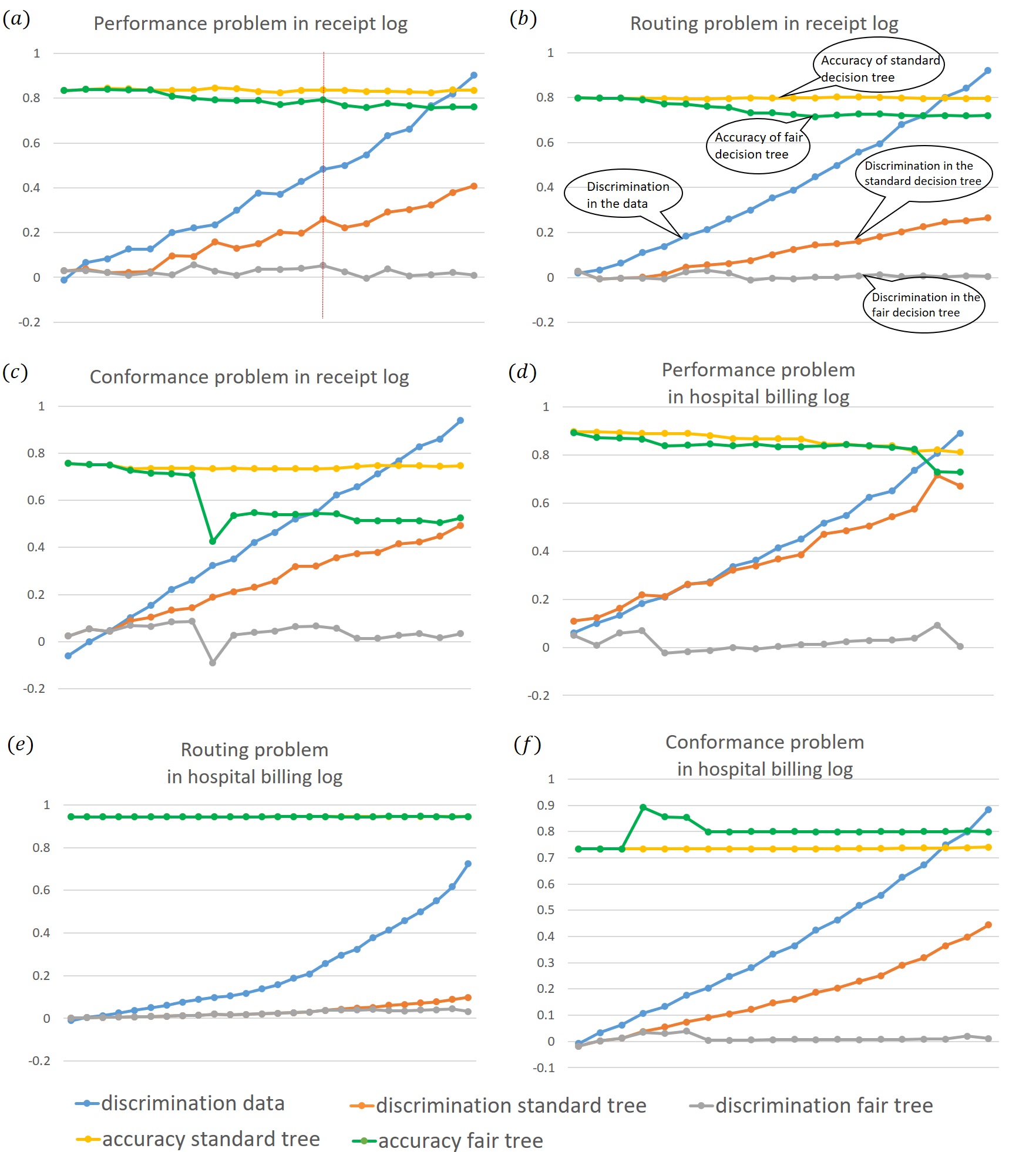}
		\caption{The result of applying the implemented ProM plug-in on two real event logs. In all these charts, the blue curve exhibits the level of discrimination in data, the orange curve shows the level of discrimination in standard decision tree, the gray curve shows the level of discrimination in a fair decision tree, the yellow color curve exhibits, and, the green color curve exhibits the accuracy of the fair tree. The first three pictures respectively show the results of applying our technique when there is (a) a performance problem, (b) a routing problem, and (c) a conformance problem in the receipt process. (d), (e) and (f) show the same results when the same problems are observed in the hospital billing process. In all these experiments $\epsilon =0.05$ and the level of discrimination in the data is depicted by the $x$-axis. In all these experiments, the level of discrimination in the fair classifiers are less than the given threshold $\epsilon$. Also, as the level of discrimination increases in the data, the difference between the accuracy of the fair decision tree and standard decision tree increases. Also in part (f), the fair decision tree demonstrates a better performance than the standard decision tree in terms of accuracy.}\label{pic:resultall}
	\end{figure}
	\begin{SCfigure}
		%	\floatbox[{\capbeside\thisfloatsetup{capbesideposition={right,top},capbesidewidth=4cm}}]{figure}[\FBwidth]
		\caption{The result of applying implemented plug-in with different values for parameter $\epsilon$ which is depicted in purple in the chart. In this chart, the value of $\epsilon$ shown by the pink curve. The level of discrimination in the data in all these experiments are the same. In all these experiments, the level of discrimination in the fair decision tree is lower than the given threshold $\epsilon$. Also, the accuracy of the fair decision tree tends to be lower for the lower values of $\epsilon$.}\label{pic:epsilon}
		\includegraphics[width=6.5cm]{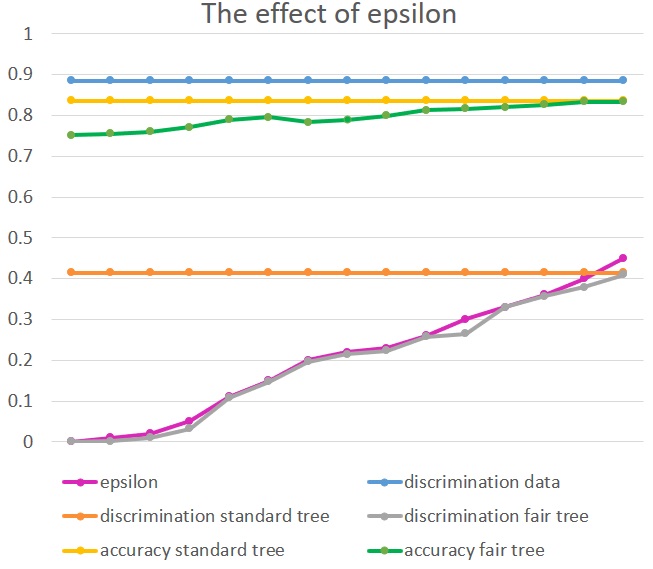}
	\end{SCfigure}	
	The results of our experiment are depicted in Figure~\ref{pic:resultall}. The first three charts present the results for the receipt process and the other three charts present the results for the hospital billing process regarding the three problems that have been mentioned above. Chart (a) and chart (d) show the results of applying our technique when there is a performance problem in the process for which we consider the delay in the traces. Chart (b) and chart (e) show the results of applying our technique when there is a routing problem in the process for which we consider the choice between ``T05 Print and send confirmation of receipt" and skipping this transition in the receipt process and the choice between ``BILLED" and skipping this transition in the hospital billing process. Chart (e) and chart (f) show the results of applying our technique when there is a conformance problem in the process for which we consider the existence of deviation in the traces in both processes.
	
	In each chart, the level of discrimination in the data is depicted on the $x$-axis (and also by the blue curve for the sake of easy comparison). In each experiment the level of discrimination in data (denoted by `discrimination data', the blue curve, in the charts), the level of discrimination in standard decision tree (denoted by `discrimination standard tree', the orange curve, in the charts), the level of discrimination in fair decision tree (denoted by `discrimination fair tree', the gray curve, in the charts), the accuracy of the standard tree (denoted by `accuracy standard tree', the yellow color curve, in the charts), and, the accuracy of the fair tree (denoted by `accuracy fair tree', the green color curve, in the charts) are recorded. For example the two trees shown in Figure \ref{pic::snapshot} are related to the red vertical line in Figure \ref{pic:resultall}(a), where the discrimination in the data is 0.48, the discrimination by the standard decision tree is 0.26, the discrimination by the fair decision tree is 0.05, and the accuracy of standard and fair decision tree are 0.84 and 0.79, respectively ($\epsilon = 0.05$ and we set no limit for applying relabeling technique).
	
	As is depicted in Figure~\ref{pic:resultall}, we can reduce the discrimination on the sensitive group at the expense of some reduction at the accuracy of the classifier. As expected, as the level of discrimination increases in the data, the amount of the accuracy of the classifier that needs to be sacrificed for removing the discrimination increases. We need to be careful using this technique, as there are occasions where discrimination may be put on the favorable group. This phenomenon is also unfair. Surprisingly, in some cases like Figure~\ref{pic:resultall}(f), the fair decision tree outperforms the standard decision tree in terms of accuracy. This phenomenon has been reported in~\cite{kamiran2010discrimination} as well. Note that in all the experiments the relabeling technique with no limitation on the label of the leaf have been used, which means, both leaves with + and - labels may get relabeled. 

	The chart in Figure~\ref{pic:epsilon}, demonstrates the level of discrimination in the fair decision tree and its accuracy for different settings of parameter $\epsilon$. As expected, the accuracy of the fair decision tree is lower when $\epsilon$ is smaller. Here, we use the receipt log and delay in traces as the class situation feature.
	
	To demonstrate that a fair classifier may affect the results of the root cause analysis of a given problem in a process, an example based on an experiment on a real event log is given in the Appendix \ref{appendix}.
		%---------------------------------------------------------------
	\section{Conclusion}\label{conclution}
	%---------------------------------------------------------------	
	The first step toward enhancing a process by removing one of its performance or compliance problems is diagnosing the root causes of that problem. By using standard data mining techniques for detecting the causes, the results might be obvious and mainly regarding those parts of the process that can not be altered. The other less obvious roots of the given problem are sometimes the most precious diagnosis where the real improvement can be applied. To reveal these less vivid causes of the problem we need to mask the obvious ones. We did so by looking at the cause that we need to ignore its effect on the problem as the sensitive attribute. Then removing the dependency between the sensitive and the class attributes from the created classifier, we allow other causes to shine. This is done at the expense of a small reduction in the accuracy of the resulting classifier.
	
	This research has several applications; detecting the discrimination within a process, removing the discrimination from the process by replacing the fair classifier with the current one, making more accurate and realistic judgments about the root causes of the problem at hand.
	
	This research can be extended in several directions. The first one is to add new derived attributes to the log when enriching the log. The other one is altering the fairness criteria, the classification method, or the technique for creating the discrimination-free classifier depending on the application.

	%---------------------------- Bibliography -------------------------------
	
	\bibliographystyle{splncs04}
	\bibliography{biblo2}

\begin{thebibliography}{10}
\providecommand{\url}[1]{\texttt{#1}}
\providecommand{\urlprefix}{URL }
\providecommand{\doi}[1]{https://doi.org/#1}

\bibitem{aghaei2019learning}
Aghaei, S., Azizi, M.J., Vayanos, P.: Learning optimal and fair decision trees
  for non-discriminative decision-making. arXiv preprint arXiv:1903.10598
  (2019)

\bibitem{berk2018fairness}
Berk, R., Heidari, H., Jabbari, S., Kearns, M., Roth, A.: Fairness in criminal
  justice risk assessments: The state of the art. Sociological Methods \&
  Research p. 0049124118782533 (2018)

\bibitem{bolukbasi2016man}
Bolukbasi, T., Chang, K.W., Zou, J., Saligrama, V., Kalai, A.: Man is to
  computer programmer as woman is to homemaker? debiasing word embeddings. In:
  Proceedings of the 30th International Conference on Neural Information
  Processing Systems. pp. 4356--4364. NIPS'16, Curran Associates Inc., USA
  (2016), \url{http://dl.acm.org/citation.cfm?id=3157382.3157584}

\bibitem{dwork2012fairness}
Dwork, C., Hardt, M., Pitassi, T., Reingold, O., Zemel, R.: Fairness through
  awareness. In: Proceedings of the 3rd Innovations in Theoretical Computer
  Science Conference. pp. 214--226. ITCS '12, ACM, New York, NY, USA (2012).
  \doi{10.1145/2090236.2090255},
  \url{http://doi.acm.org/10.1145/2090236.2090255}

\bibitem{sani2017subgroup}
{Fani Sani}, M., {van der Aalst}, W., {Bolt Irondo}, A., Garc{\'i}a-Algarra,
  J.: Subgroup discovery in process mining. In: Abramowicz, W. (ed.) Business
  Information Systems. pp. 237--252. Lecture Notes in Business Information
  Processing, Springer, Germany (2017). \doi{10.1007/978-3-319-59336-4\_17}

\bibitem{hardt2016equality}
Hardt, M., Price, E., Srebro, N.: Equality of opportunity in supervised
  learning. In: Proceedings of the 30th International Conference on Neural
  Information Processing Systems. pp. 3323--3331. NIPS'16, Curran Associates
  Inc., USA (2016), \url{http://dl.acm.org/citation.cfm?id=3157382.3157469}

\bibitem{Hardt:2016:EOS:3157382.3157469}
Hardt, M., Price, E., Srebro, N.: Equality of opportunity in supervised
  learning. In: Proceedings of the 30th International Conference on Neural
  Information Processing Systems. pp. 3323--3331. NIPS'16, Curran Associates
  Inc., USA (2016), \url{http://dl.acm.org/citation.cfm?id=3157382.3157469}

\bibitem{kamiran2010classification}
Kamiran, F., Calders, T.: Classification with no discrimination by preferential
  sampling. In: Informal proceedings of the 19th Annual Machine Learning
  Conference of Belgium and The Netherlands (Benelearn'10, Leuven, Belgium, May
  27-28, 2010). pp.~1--6 (2010)

\bibitem{kamiran2012data}
Kamiran, F., Calders, T.: Data preprocessing techniques for classification
  without discrimination. Knowledge and Information Systems  \textbf{33}(1),
  1--33 (Oct 2012). \doi{10.1007/s10115-011-0463-8},
  \url{https://doi.org/10.1007/s10115-011-0463-8}

\bibitem{kamiran2010discrimination}
Kamiran, F., Calders, T., Pechenizkiy, M.: Discrimination aware decision tree
  learning. In: Proceedings of the 2010 IEEE International Conference on Data
  Mining. pp. 869--874. ICDM '10, IEEE Computer Society, Washington, DC, USA
  (2010). \doi{10.1109/ICDM.2010.50},
  \url{http://dx.doi.org/10.1109/ICDM.2010.50}

\bibitem{kleinberg2016inherent}
Kleinberg, J.: Inherent trade-offs in algorithmic fairness. SIGMETRICS Perform.
  Eval. Rev.  \textbf{46}(1),  40--40 (Jun 2018).
  \doi{10.1145/3292040.3219634},
  \url{http://doi.acm.org/10.1145/3292040.3219634}

\bibitem{kusner2017counterfactual}
Kusner, M.J., Loftus, J., Russell, C., Silva, R.: Counterfactual fairness. In:
  Guyon, I., Luxburg, U.V., Bengio, S., Wallach, H., Fergus, R., Vishwanathan,
  S., Garnett, R. (eds.) Advances in Neural Information Processing Systems 30.
  pp. 4066--4076. Curran Associates, Inc. (2017),
  \url{http://papers.nips.cc/paper/6995-counterfactual-fairness.pdf}

\bibitem{leemans2014process}
Leemans, S., Fahland, D., {Aalst, van der}, W.: Process and deviation
  exploration with inductive visual miner pp. 46--50 (2014)

\bibitem{de2016general}
de~Leoni, M., van~der Aalst, W.M., Dees, M.: A general process mining framework
  for correlating, predicting and clustering dynamic behavior based on event
  logs. Inf. Syst.  \textbf{56}(C),  235--257 (Mar 2016).
  \doi{10.1016/j.is.2015.07.003},
  \url{https://doi.org/10.1016/j.is.2015.07.003}

\bibitem{RafieiWA18}
Rafiei, M., von Waldthausen, L., van~der Aalst, W.M.P.: Ensuring
  confidentiality in process mining. In: Proceedings of the 8th International
  Symposium on Data-driven Process Discovery and Analysis {(SIMPDA} 2018),
  Seville, Spain, December 13-14, 2018. pp. 3--17 (2018),
  \url{http://ceur-ws.org/Vol-2270/paper1.pdf}

\bibitem{rozinat2006decision}
Rozinat, A., van~der Aalst, W.M.P.: Decision mining in {P}ro{M}. In: Dustdar,
  S., Fiadeiro, J.L., Sheth, A.P. (eds.) Business Process Management. pp.
  420--425. Springer Berlin Heidelberg, Berlin, Heidelberg (2006)

\bibitem{zafar2017fairness}
Zafar, M.B., Valera, I., Gomez~Rodriguez, M., Gummadi, K.P.: Fairness beyond
  disparate treatment \& disparate impact: Learning classification without
  disparate mistreatment. In: Proceedings of the 26th International Conference
  on World Wide Web. pp. 1171--1180. WWW '17, International World Wide Web
  Conferences Steering Committee, Republic and Canton of Geneva, Switzerland
  (2017). \doi{10.1145/3038912.3052660},
  \url{https://doi.org/10.1145/3038912.3052660}

\bibitem{zemel2013learning}
Zemel, R., Wu, Y., Swersky, K., Pitassi, T., Dwork, C.: Learning fair
  representations. In: Proceedings of the 30th International Conference on
  International Conference on Machine Learning - Volume 28. pp.
  III--325--III--333. ICML'13, JMLR.org (2013),
  \url{http://dl.acm.org/citation.cfm?id=3042817.3042973}

\bibitem{zliobaite2015survey}
Zliobaite, I.: A survey on measuring indirect discrimination in machine
  learning. arXiv preprint arXiv:1511.00148  (2015)

\end{thebibliography}
	%---------------------------- Appendix -------------------------------
%	\begin{appendices}
	\appendix
	\section{Example}   \label{appendix}
\begin{figure}
		\includegraphics[width=120mm]{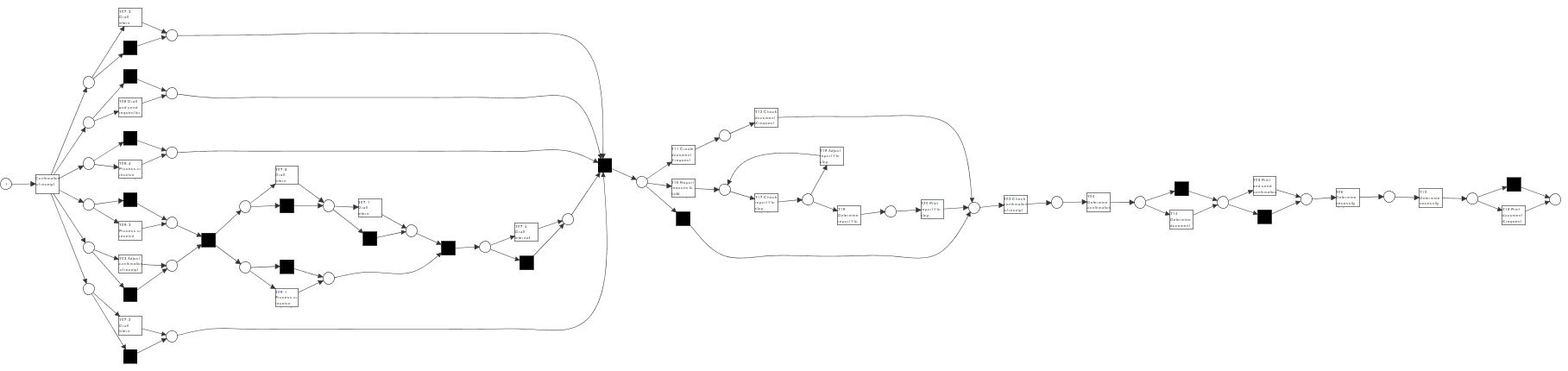}
		\caption{The Petri-net of the receipt process generated by Inductive Miner.}\label{pic:recieptPN}
	\end{figure}

	To demonstrate how creating a fair classifier may affect the results of the root cause analysis in a given process, consider the receipt process log (This is the same event log that has been used in Section.~\ref{result}). This event log contains the records of the execution of the receiving phase of the building permit application process in an anonymous municipality. A Petri-net model for this process is depicted in Figure \ref{pic:recieptPN}. One of the trace attributes in this log is \emph{responsible} which may takes one of 38 different values indicating which resource is responsible for the corresponding case. Note that the resource responsible for the trace may be different from the one who executes the trace. It is known that some of the resources are more busy than others. Also, it is observed that in the given process some of the traces have been delayed. We can augment the log with the \emph{trace-delay} attribute ($\sitf_{trace,delay}$) where the threshold for the delay is set to 2 percent of the maximum duration of all traces in this log (note that in this event log the average duration of traces is 467263915 milliseconds which is roughly 1.96 percent of the maximum duration of a trace in this event log); so, $values(\sitf_{trace,delay})) =\{on$-$ time, delayed\}$. In this log, we consider $\sitf_{trace,responsible}$ as the sensitive attribute and the overloaded resources as the sensitive values, also, $\sitf_{trace,delay}$ as the class attribute and $on$-$time$ as the desirable outcome. The considered situation specification is as follow:
	\begin{itemize}
		\item $\EP =\{ \sitf_{T06\ Determine\ necessity\ of\ stop\ advice,Resource}, \\\sitf_{T06\ Determine\ necessity\ of\ stop\ advice,Resourc\ Workload}, \sitf_{trace, responsible},\\\sitf_{trace, channel},\sitf_{trace, department},\sitf_{trace,group},\sitf_{trace, deadline},\sitf_{trace, deviation},\\\sitf_{trace, number\ modelMove},\sitf_{trace, number\ logMove},\}$.
		\item $\ssf = \sitf_{trace, responsible}$, where the set of sensitive values of this situation feature is $$\{Resource07,Resource08,Resource09,Resource11,Resource15,$$
		$$Resource17,Resource18\}.$$
		\item $\csf = trace$-$delay$, where $values(trace$-$delay)=\{on$-$time,delayed\}$ which are renamed as $values(trace$-$delay)=\{desirable\ class,undesirable\ class\}$.
		\item $\epsilon = 0$.
	\end{itemize}

	With this setting, the level of discrimination in this data log is 0.0954. Even though we did not use te values of $\sitf_{trace,responsible}$ for the creation of the decision tree, the discrimination caused by the standard decision tree is 0.126 which means that this tree could detect and amplify the discrimination through the correlation among the different situation feature values. 
	
		\begin{figure}
		\includegraphics[width=120mm]{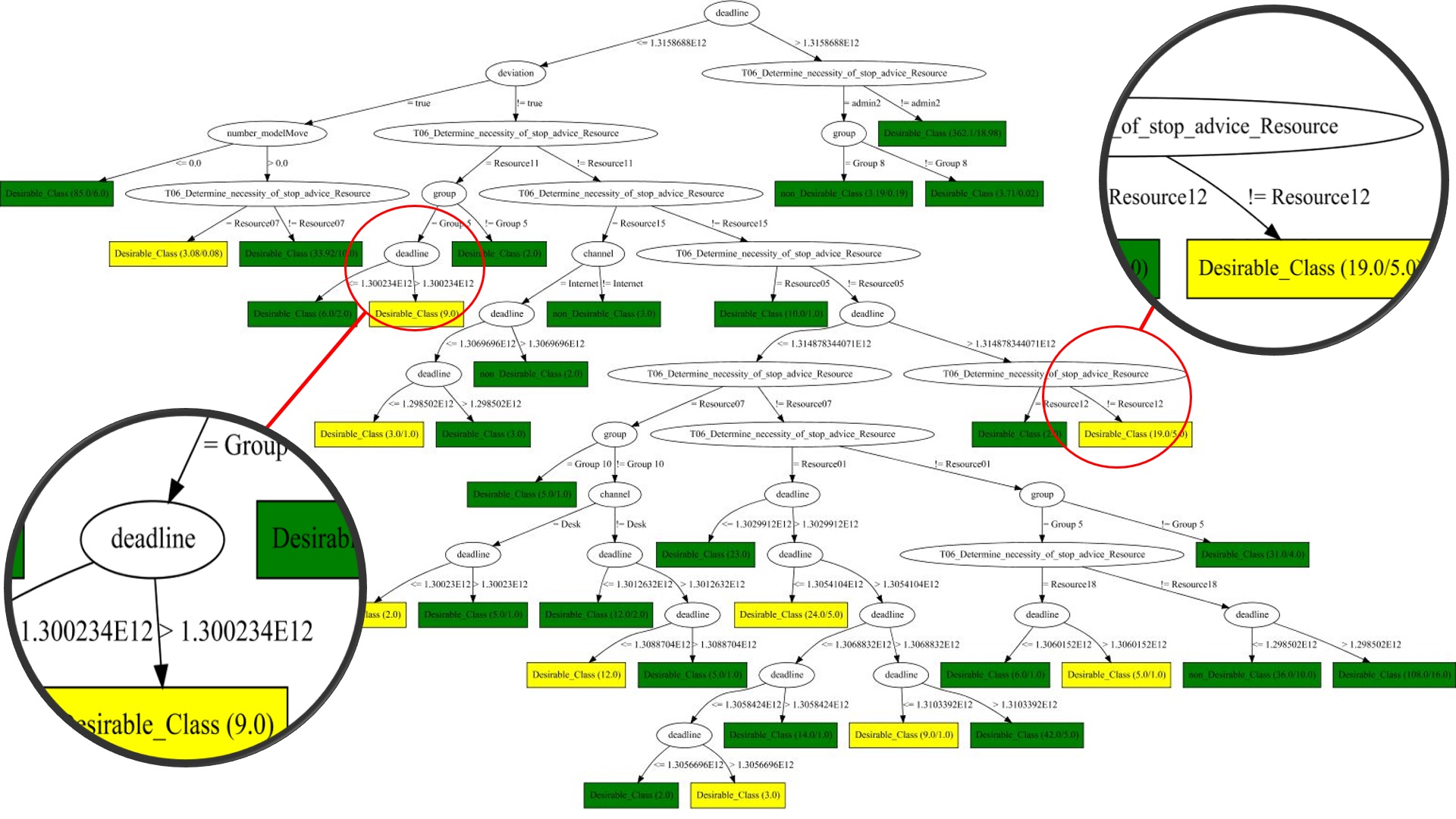}
		\caption{The Fair decision tree created by applying the implemented plug-in on the receipt process. The relabeled leaf in the left side of the tree includes 9 situations, all sensitive and delayed. The relabeled leaf on the right side of the tree includes 24 situations, 12 sensitive which 10 of them were delayed and 7 favorable situations which 4 of them were delayed.}\label{pic:recieptFairTree}
	\end{figure}

	To remove the correlation among the $\sitf_{trace,responsible}$ and $\sitf_{trace,delay}$, we create a fair tree in which $\epsilon =0$. Note that we can convey that the delay caused by the busier resources are explainable to some extend (but not all the delay caused by them) by setting $\epsilon$ to a value higher than zero. To have meaningful results, we restrict using the relabeling for leaves with undesirable outcome, i.e. with the \emph{delayed} label. The resulting fair decision tree is shown in Figure \ref{pic:recieptFairTree}. Differently from the standard decision tree, after removing the discrimination, ten classes of situations are not considered problematic in the fair decision tree. For example, those situations in which $$\sitf_{trace,deadline}\leq 1.3158688E12\ \land\ \sitf_{trace,group} = Group5\ \land$$ $$\sitf_{T06\ Determine\ necessity\ of\ stop\ advice,Resource}=Resource11\ \land$$ $$\sitf_{trace,deviation}=true\ \land \  \sitf_{trace,deadline} > 1.300234E12$$ are considered $on-time$. This rule applies to 9 situations, all were sensitive and delayed. Here, those leaves are relabeled that relabeling them have the minimum negative impact on the correctness of the decision tree while removing the dependency between the sensitive and class situation features. Using the relabeling technique, the leaves with ``delayed'' label which have just or mainly sensitive instances are more likely to be relabeled. This way the rules that apply specifically to the sensitive instances are ignored. The remaining inferred rules from the fair decision tree that apply for the sensitive instances are those rules that have mainly favorable situations with undesirable label. We can interpret these rules as the ones targeting those sensitive situations which would have been delayed even if they had been favorable. These leaves may signify those causes of the problem that are general and not related to the situation being sensitive and might be improved. 
	
	It is worth noting that if we just remove the instances corresponding to those situations which belong to the sensitive group and were delayed from the data, the resulting decision tree would be a decision tree with three leaves which is not informative.

%\end{appendices}	

	%\section{Detailed proof of theorem \ref{NEk2SRG}}\label{APB}
	
\end{document}